\documentclass[12pt,epsfig]{article}

\usepackage{amsmath}
\usepackage{graphicx}
\usepackage[psamsfonts]{amssymb}
\usepackage{cite}
\numberwithin{equation}{section}

\newlength{\dinwidth}
\newlength{\dinmargin}
\newcommand{\ba}{\begin{array}}
\newcommand{\ea}{\end{array}}
\newcommand{\be}{\begin{eqnarray}}
\newcommand{\ee}{\end{eqnarray}}

\def\vectt[#1,#2]{\left(%
\begin{array}{c} #1 \\ #2 \end{array} \right)}

\def\trivectt[#1,#2,#3]{\left(%
\begin{array}{c} #1 \\ #2 \\ #3 \end{array} \right)}

\def\nn{\nonumber \\}

\def\d{{\rm d}}

\newcommand{\gsim}{\mathrel{\mathop{\kern 0pt \rlap
  {\raise.2ex\hbox{$>$}}}
  \lower.9ex\hbox{\kern-.190em $\sim$}}}
\newcommand{\lsim}{\mathrel{\mathop{\kern 0pt \rlap
  {\raise.2ex\hbox{$<$}}}
  \lower.9ex\hbox{\kern-.190em $\sim$}}}

\begin{document}
\thispagestyle{empty} \addtocounter{page}{-1}
\begin{flushright}
SNUST 030601\\
{\tt hep-th/0306148}\\
\end{flushright} \vspace*{1cm}
\centerline{\Large \bf  Can Branes Travel Beyond CTC Horizon}
\vspace*{0.3cm} \centerline{\Large \bf in}
\vspace*{0.3cm} \centerline{\Large \bf G\"odel
Universe?~\footnote{Work supported in part by the KOSEF
Interdisciplinary Research Grant 98-07-02-07-01-5 and the KOSEF
Leading Scientist Grant.}} \vspace*{1.5cm}
\centerline{\bf Yasuaki Hikida ${}^a$ {\rm and} Soo-Jong Rey
${}^{a,b}$} \vspace*{1.0cm}
\vskip0.5cm
\centerline{\it School of Physics \& BK-21 Physics Division}
\vspace*{0.25cm}
\centerline{\it Seoul National University, Seoul 151-747 \rm KOREA
${}^{a}$}
\vspace*{0.5cm}
\centerline{\it School of Natural Sciences, Institute for Advanced
Study} \vspace*{0.25cm}
\centerline{\it Einstein Drive, Princeton NJ 08540 \rm USA ${}^b$}
\vspace*{0.8cm}
\centerline{\tt hikida@phya.snu.ac.kr \hskip1cm
sjrey@gravity.snu.ac.kr}
\vskip1.5cm
\centerline{\bf abstract}
\vspace*{0.5cm}
G\"odel universe in M-theory is a supersymmetric and homogeneous
background with rotation and four-form magnetic flux. It is known
that, as seen in inertial frame of co-moving observer, all
geodesics with zero orbital angular momentum orbit inside `surface
of light velocity' (CTC horizon). To learn if other probes can
travel beyond the CTC horizon, we study dynamics of M-graviton
and, in particular, M2-brane, whose motion is affected by Lorentz
force exerted by the four-form magnetic flux and by nonzero
orbital angular momentum. Classically, we find that both probes
gyrate closed orbits, but diameter and center of gyration depends
on sign and magnitude of probe's energy, charge and orbital
angular momentum. For M2-brane, orbits in general travel outside
the CTC horizon. Quantum-mechanically, we find that wave function
and excitation energy levels are all self-similar. We draw analogy
of probe's dynamics with Landau problem for charged particle in
magnetic field.

%\vspace*{1.1cm}

%\centerline{Submitted to Nuclear Physics B}
\baselineskip=18pt
\newpage
%%%%%%%%%%%%%%%%%%%%%%%%%%%%%%%%%%%%%%%%%%%%%%%%%%%%%%%%%%%%%%%%%%%%%%
\rightline{$\bullet$ \quad Albert Einstein \qquad ----- \qquad \sl
For us believing physicists, } \rightline{\sl the distinction
between the past, the present, and the future is only an illusion.
}
\vskip0.5cm
\section{Introduction}
The G\"odel universe \cite{godel} is a homogeneous space-time with
pressureless matter and negative cosmological constant, featured
by rotation relative to the local inertial frame associated with
each co-moving observer, and, as a result of it, no "absolute"
time function \footnote{Technically, a "time function" is a
smooth, differentiable scalar function $t(x)$ such that
$(\partial_m t)(\partial^m t) > 0$ everywhere in space-time.}, the
latter meaning that the space-time does not admit a foliation by
globally space-like hyper-surfaces. Because of these difficulties,
G\"odel universe has brought up many puzzling issues in Einstein's
general relativity: the space-time displays closed null curves
(CNCs) and closed time-like curves (CTCs), Cauchy problem is
ill-defined, and, for quantum field theories defined on it, no
obvious notion of unitary Hamiltonian evolution exists. These
issues are all concerned with the notion of chronology, on which
Hawking has put forward so-called "chronology protection
conjecture" (CPC) \cite{hawking}.

String theory and M-theory admit the G\"odel universe as a
super-symmetric solution \cite{solutions, harmark}, where, quite
importantly, requisite 4-form field strengths are also turned on.
It is in this setting that one might hope to gain a better
understanding of the above conceptual puzzles and shed light on an
eventual resolution. With such motivations, recently, several
works have revisited the issue of chronology and (stringy version
of) CPC \cite{gibbons, herdeiro, dyson, leigh, Goedel}. Among
them, Boyda et.al. \cite{Goedel} claimed that the chronology is
well definable once the G\"odel universe is prescribed with the
macroscopic holography \cite{bakrey, bousso}. In effect, their
proposal argues for keeping (a part of) the causal region
associated with a co-moving observer and replacing the rest (which
includes the CTC region) by an observer-dependent holographic
screen. Most recently, Drukker et.al. \cite{drukker} have studied
dynamics of BPS super-tube (cylindrical D2-brane supported by
fundamental string and D0-brane charges) in Type IIA G\"odel
universe, and claimed that the super-tube develops an instability
in the CTC region despite being a BPS configuration.

To draw some insight concerning the CTCs in the G\"odel universe,
in this work, we introduce a probe M-graviton and M2-brane and
study their dynamics. Geodesics in the G\"odel universe was
studied previously \cite{chandrasekhar, Goedel}. There, it was
noted that null and time-like geodesic motions trace gyration
orbits whose diameter is set by the energy, and most notably, none
of these geodesics never get into the CTC region. An
interpretation based on the result, which seems to be implicit in
the holographic proposal of \cite{Goedel}, may be that the CTC
region, though it exists, is never reachable by a stationary
co-moving observer.

An immediate question is whether the feature that geodesic orbits
never get into the CTC region holds also for all other available
probes in M-theory. To answer this question, in this work, we
study classical and quantum dynamics of a probe M-graviton and, in
particular, M2-brane in the G\"odel universe. The reason why
M2-brane might serve as a useful probe for the question posed is
because of the following intuitive argument. In addition to the
rotating space-time metric, the G\"odel universe in M-theory is
characterized by constant 4-form magnetic field strengths, whose
strength is set by the strength of the rotation. M2-brane is an
electrically charged object coupled minimally to the 4-form field,
so it would experience the 4-form Lorentz force in addition to the
geodesic force acted upon by the G\"odel metric. This means that,
under suitable circumstances (which we will spell out explicitly
later), the Lorentz force might well cancel the geodesic force. In
that case, M2-brane's orbit can be larger than the null geodesics,
and would eventually be able to travel into the region where
co-moving observers perceive CTCs present.

We find it useful to compare the situation with the well-known
Landau problem: dynamics of a charged particle on a homogeneous
2-space (such as sphere ${\bf S}^2$, plane ${\bf R}^2$, and
pseudo-sphere ${\bf H}^2$) under a uniform magnetic field. Take,
for definiteness, the BPS particle $m = |q|$ on a plane ${\bf
R}^2$. On ${\bf R}^2$, geodesics are straight lines. The magnetic
field exerts Lorentz force to the BPS particle, causing it to
undergo gyration around the Larmor orbit, whose radius of
curvature is set by initial velocity ${\bf v}$ as
\be R_{\rm L} = \left({c \over B}\right) \vert {\bf v} \vert.
\label{larmorradius} \ee
Quantum mechanically, the particle dynamics is describable as that
of two-dimensional simple harmonic oscillator, whose energy
spectrum is given in terms of non-negative quantum number ${\bf
n}$ as
\be E_{\rm L} = \Omega_{\rm L} \left({\bf n} + {1 \over 2} \right)
\qquad {\rm where} \qquad \Omega_L = \left({B \over c}\right).
\label{larmorfreq} \ee
As ${\bf R}^2$ is a homogeneous space, the gyration takes place
isomorphically everywhere and its center can be brought to, say,
the center $O$ by Killing transformations. Dynamics on ${\bf S}^2$
or ${\bf H}^2$ is further corrected by curvature-dependent
contribution $\Delta E_{\rm L} = \pm {1 \over 2 R^2} ({\bf n} + {1
\over 2})^2$ to the energy spectrum Eq.(\ref{larmorfreq}), where
$R$ is the radius of curvature, but otherwise qualitatively the
same. In particular, gyration orbit is closed for strong magnetic
field and its center can be brought to to the center ${\cal O}$ by
Killing transformations.

This paper is organized as follows. In section 2, we investigate
classical dynamics of a probe brane in one of the simpler G\"odel
universes in M-theory, ${\cal G}_5 \times X_6$. We study both
M-graviton and M2-brane probes, and show that M2-brane orbit can
extend beyond the CTC horizon. In section 3, we study quantum
dynamics of these probe branes. We find that the energy spectrum
is discrete and is strikingly reminiscent of that of the Landau
problem, Eq.(\ref{larmorfreq}). Section 4 is devoted to discussion
of various points worthy of further investigation.

\section{Classical Dynamics}
\subsection{Set-Up}
The G\"odel universe in M-theory is a family of classical vacua of
the form ${\cal G}_{2n+1} \times X_{10-2n}$ \cite{harmark}, where
the index $n=1, \cdots, 5$ refers to the number of rotating
planes. The simplest situation in which multiple planes are
rotating simultaneously is ${\cal G}_5 \times X_6$, and we shall
focus our consideration mainly to this case in this work
\footnote{The simplest G\"odel universe, ${\cal G}_3 \times X_8$,
involving a {\sl single} rotating plane preserves 8
supersymmetries only, as contrasted to those involving multiple
rotating planes, which preserve 16+4 = 20 supersymmetries
($n=2,3,4$) or 16+2 = 18 supersymmetries ($n=5$).}. The metric of
${\cal G}_5 \times X_6$ is given by
\be \d s^2_{\rm M} &=& - (\d t + \mu \omega )^2 + \d s^2 ({\bf
R}^4) + \d s^2 (X_6) \nn G_4 &\equiv&  \d C_3 = 2 \mu \, J \wedge
K, \label{godel}\ee
where $\omega$ is the twist one-form on the spatial slice ${\bf
R}^4$ in ${\cal G}_5$, while $J, K$ are K\"ahler two-forms on
${\bf R}^4$ and a six-dimensional `internal' manifold $X_6$,
respectively. Locally, the one-form $\omega$ is related to $J$ as
$J = \d \omega$. The constant $\mu$ is a parameter for the
simultaneous rotation of ${\bf R}^2 \oplus {\bf R}^2 \subset {\bf
R}^4$ as well as for the 4-form magnetic flux $G_4 \equiv \d C_3$.
It can be set to unity by rescaling proper distance on ${\bf R}^4$
by $1/|\mu|$, but, for the consideration of dimensional analysis
and (pseudo)-symmetry transformation, we find it useful to retain
it explicitly.

We introduce a probe M2- or $\overline{\rm M2}-$brane, which is
coupled minimally to the 3-form potential $C_3$, and examine
classical dynamics of it in the background Eq.(\ref{godel}).
Classical dynamics of the probe brane is governed by the
Born-Infeld action
\be S = -T_2 \int_{\Sigma_3} \sqrt{ - \det X^* g} - Q
\int_{\Sigma_3} X^* C_3\, \label{m2action} \ee
where $X^*g, X^*C_3$ are the pull-backs of the metric and the
3-form potential in Eq.(\ref{godel}) on the world-volume of
M2-brane. Noether charge of the M2-brane is denoted as $Q$,
measured in unit of the M2-brane tension $T_2$. Thus, for a BPS
M2-- or $\overline{\rm M2}$--brane, $Q = T_2$ or $Q = - T_2$,
respectively.

For definiteness, consider a M2-brane wrapped on a supersymmetric
two-cycle $\Sigma$ in $X_6$ \footnote{The analysis of
\cite{harmark} indicates that this is a supersymmetric
configuration.}. Rigid dynamics of the M2-brane is describable by
dimensionally reducing Eq.(\ref{godel}) to five-dimensional
G\"odel universe ${\cal G}_5$ threaded with the magnetic flux $G_2
\equiv \oint_\Sigma G_4$ (up to normalization):
\be \d s^2_{5} &=& -(\d t + \mu \omega)^2 + \d s^2 ({\bf R}^4) \nn
G_2 &\equiv& \d C_1 = 2 \sqrt{3} \mu \, J , \label{reducedgodel}
\ee
and the M2-brane reduces to a charged point-particle of mass $m =
T_3 {\rm vol}(\Sigma)$ and electric charge $q = Q {\rm
vol}(\Sigma)$, coupled minimally to the `vector' potential $C_1$.
From Eq.(\ref{m2action}), the effective world-line action of the
5-dimensional charged particle is given in local reparametrization
invariant way as
\be
 S = \ \int \d \tau
 \left[ \frac12 \left(e^{-1} g_{mn} (x) \dot x^{m} \dot x^{n}
        - e \, m^2 \right) - {q \over 2 \sqrt{3}} C_{1~m}(x) \dot x^{m}
        \right] ~, \quad (m,n=0, 1,\cdots, 4)
\label{action} \ee
where $\dot{x}^m \equiv \d x^m / \d \tau$. We solve the equation
of motion for $x^m(\tau)$: \be
  {D \dot{x}^m \over D \tau}
   - \frac{\dot e}{e} \dot x^m
   +  e {q \over 2 \sqrt{3}} {G^{m}_2}_{n} \dot x^{n} = 0 ~,
   \nonumber
\ee
where $D/D\tau$ refers to the covariant derivative with respect to
the Christoffel connection $\Gamma^m_{np}$ of the metric
Eq.(\ref{reducedgodel}). Equation of motion for the worldline
einbein $e(\tau)$ is
\be
 e(\tau) = m^{-1} \sqrt{-g_{mn}(x) \dot x^{m} \dot x^{n}} ~.
\label{affine} \ee
The worldline action Eq.(\ref{action}) is invariant under local
reparametrization invariance, $\tau \rightarrow f(\tau)$ and $e
\rightarrow e / |\dot{f}(\tau)|$, and we fix it by choosing the
gauge $e = +1$, so that $\tau$ is the affine parameter of the
world-line. The gauge-fixed equations of motion for $x^m(\tau)$
are
\begin{equation}
{D \dot{x}^m \over D \tau}
   + {q \over 2 \sqrt{3}} {G^{m}_2}_{~n} \dot x^{n} = 0 ~,
\label{eom}
\end{equation}
subject to the constraint Eq.(\ref{affine}).

To proceed further, we find it convenient to express the metric
Eq.(\ref{reducedgodel}) in the bipolar coordinates (see, e.g.,
Eq.(3.7) in \cite{Goedel}) as:
\be \d s^2_5 &=& - \d t^2 - 2 \mu \left(r^2_1 \d \phi_1 + r^2_2 \d
\phi_2 \right) \d t
        - 2\mu^2 r^2_1 r^2_2 \d \phi_1 \d \phi_2  \nonumber \\
       & &+ r^2_1 \left(1 - \mu^2 r^2_1 \right) \d \phi^2_1
       + r^2_2 \left(1 - \mu^2 r^2_2 \right) \d \phi^2_2 + \d r^2_1 + \d r^2_2~,
\label{metric} \ee
and the one-form connection $C_1$ in the Landau gauge as:
\be  G_2 \equiv \d C_1 = 2 \sqrt3 \mu \, \d (r_1^2 \wedge \d
\phi_1 + r_2^2 \wedge \d \phi_2). \label{landaugauge} \ee
We refer the origin of the coordinate system as $O$. Notice that
the bipolar coordinates Eq.(\ref{metric}) displays explicitly that
the signature of $\phi_{1,2}$ line element flips sign across the
hyper-surface $\sqrt{r_{1}^2 + r_{2}^2} = 1/|\mu|$. Thus, at the
hyper-surface, paths around $\phi_1$ or $\phi_2$ are CNCs, and,
beyond the hyper-surface, they are CTCs. As such, we will refer
these codimension-one hyper-surfaces located at
\be R_{\rm CTC} = {1 \over |\mu|} \label{ctcsurface} \ee
as "CTC horizon". We emphasize again that the "CTC horizon" is a
notion set forth specifically by each co-moving observer, whom we
have put conveniently at the origin $O$. As G\"odel universe is a
homogeneous space, co-moving observers placed at any location on
$G_5$ are all equivalent and can be always brought to the origin
$O$.

Using explicit form of the Christoffel connections as recollected
in the appendix, we find the equations of motion Eq.(\ref{eom})
are given by
%\
\be && \ddot t + 2 \mu^2 r_1 \dot r_1 \dot t
  %+ 2 \mu^2 r_2 \dot r_2 \dot t
  + 2 \mu^3 r_1^3 \dot \phi_1 \dot r_1
  + 2 \mu^3 r_1 r^2_2 \dot \phi_2 \dot r_1
%   + 2 \mu^3 r^3_2 \dot \phi_2 \dot r_2
%  + 2 \mu^3 r_2 r_1^2 \dot \phi_1 \dot r_2 \nonumber \\
%  & \qquad \qquad \qquad \qquad \qquad \qquad
   + 2 q \mu^2  r_1 \dot r_1 +
   %r_2 \dot r_2
   \left( 1 \leftrightarrow 2 \right) = 0~, \label{t-eom} \\
   \nonumber \\
  && \ddot r_1 + 2 \mu r_1 \dot \phi_1 \dot t
    - r_1 (1 - 2 \mu^2 r^2_1) \dot \phi_1^2
    + 2 \mu^2 r_1 r_2^2 \dot \phi_1 \dot \phi_2
    + 2 q \mu  r_1 \dot \phi_1 = 0 ~,
      \label{r_1eom} \\
      \nonumber \\
%  &\ddot r_2 + 2 \mu r_2 \dot \phi_2 \dot t
%    - r_2 (1 - 2 \mu^2 r^2_2) \dot \phi_2 \dot \phi_2
%    + 2 \mu^2 r_2 r_1^2 \dot \phi_2 \dot \phi_1
%    -  2 \mu q r_2 \dot \phi_2 = 0 ~,
%      \label{r_2eom} \\
  && \ddot \phi_1 - \frac{2 \mu}{r_1} \dot r_1 \dot t
    + 2\left(\frac{1}{r_1} -  \mu^2 r_1 \right) \dot \phi_1 \dot r_1
    - \frac{2 \mu^2 r^2_2}{r_1} \dot \phi_2 \dot r_1
    - 2 q \mu \frac{\dot r_1}{r_1} = 0~, \label{ang-eom}
%  &\ddot \phi_2 - \frac{2 \mu}{r_2} \dot r_2 \dot t
%    + 2\left(\frac{1}{r_2} -  \mu^2 r_2 \right) \dot \phi_2 \dot r_2
%    - \frac{2 \mu^2 r^2_1}{r_2} \dot \phi_1 \dot r_2
%    + 2 \mu q  \frac{\dot r_2}{r_2} = 0~,
\ee and similar ones with $(1 \leftrightarrow 2)$ for the latter
two equations of motion.

The G\"odel universe Eq.(\ref{reducedgodel}) or, equivalently,
Eqs.(\ref{metric}, \ref{landaugauge}) exhibits the following
symmetries:
\be {\rm P} \quad : \quad && \qquad \omega \rightarrow - \omega
\qquad {\rm and} \qquad \mu \rightarrow - \mu \label{muparity} \\
{\rm PT} \quad : \quad && t \rightarrow - t, \quad \mu \omega
\rightarrow - \mu \omega \quad {\rm and} \quad q \rightarrow - q
\label{cpt} \ . \ee
Notice that $\omega \rightarrow - \omega$ in Eq.(\ref{godel})
amounts in bipolar coordinates to parity inversion, $\phi_{1,2}
\rightarrow - \phi_{1,2}$, in each ${\bf R}^2$-plane. Thus,
Eq.(\ref{muparity}) acts as parity inversion accompanied by
inversion of the rigid rotation, while Eq.(\ref{cpt}) does as
simultaneous action of the time-reversal and parity-inversion on
${\cal G}_5$ and $X_6$. We will find later these
(pseudo)symmetries serve useful for understanding classical and
quantum dynamics of the probe objects.

As the metric and the magnetic field strength on ${\cal G}_5$ are
functions of $r_1, r_2$ only, $-\partial_t$, $\partial_{\phi_1}$,
$\partial_{\phi_2}$ are Killing vectors. Accordingly, the
canonical momenta conjugate to $-t$, $\phi_1$ and $\phi_2$ are
conserved, first integrals of motion. They are
\be
  E &=& \dot t + \mu (r_1^2 \dot \phi_1 +r^2_2 \dot \phi_2) ~,\label{E}\\
  L_1 &=& r^2_1 \dot \phi_1
  - \mu^2 r_1^2 (r_1^2 \dot \phi_1 + r^2_2 \dot \phi_2) - \mu r^2_1 (\dot t
   + q) ~,
   % \mu r^2_1 ~,
  \nonumber \\
  L_2 &=& r^2_2 \dot \phi_2
  -\mu^2 r_2^2( r^2_1 \dot \phi_1 + r^2_2\dot \phi_2)
  - \mu r^2_2 (\dot t + q) ~.
%   + 2 \sqrt3 e \beta r^2_2 ~.
\nonumber \ee
Inverting these relations,
\be \dot \phi_1
  &=& \frac{L_1}{r_1^2} + \mu (E+q) ~,
    \label{dotphi1}\\
 \dot \phi_2
  &=& \frac{L_2}{r_2^2} + \mu (E+q) ~,
    \label{dotphi2}\\
 \dot t \,\, &=& E - \mu^2 (r^2_1 + r_2^2) (E+q) - \mu (L_1 +
 L_2).
    \label{dott}
\ee
In these equations, the first terms are standard. The second term
in Eq.(\ref{dott}) reflect frame dragging rotation of the G\"odel
universe. The last term in Eq.(\ref{dott}), depending on $L_1,
L_2$, is a sort of `spin-orbit' coupling between frame-dragging
rotation and probe's orbital angular momenta.

Recall that, using homogeneity of ${\cal G}_5$, any location can
be brought to the origin $O$ ($r_1 = r_2 = 0$ in the bipolar
coordinates Eq.(\ref{metric})  by a sequence of Killing
transformations. Thus, we shall be considering a co-moving
observer located at the origin ${O}$, and let the observer perform
experiments for exploring causal structure of the G\"odel universe
by sending off the probe objects available such as M-graviton or
M2-branes. We will study first the probes with zero angular
momenta, and then those with nonzero angular momenta. In both
cases, wherever relevant, we draw close analogy with the
well-known results of the Landau problem Eqs.(\ref{larmorradius},
\ref{larmorfreq}).

\subsection{Probes with Zero Angular Momenta}
We will first consider $L_1 = L_2 = 0$ case. In this case, there
is no centrifugal barrier, and the probe brane passes through the
origin, where the co-moving observer is located. We then obtain
the radial equations of motion as exerting harmonic oscillations
isomorphically on each ${\bf R}^2$-plane:
\be
 \ddot r_1 + \mu^2 (E + q)^2 r_1= 0  \qquad {\rm and} \qquad
 \ddot r_2 + \mu^2 (E + q)^2 r_2= 0  \nonumber
\ee
subject to the gauge-fixing constraint, coupling the motion on the
two ${\bf R}^2$-planes,
\be \left\{\dot r_1^2 + (E+q)^2 \mu^2 r_1^2 \right\} + \left\{
\dot r_2^2 + (E+q)^2 \mu^2 r_2^2 \right\} = (E^2 - m^2).
\label{1stintegral} \ee
\begin{figure}
\vskip-0.5cm \centerline{\includegraphics{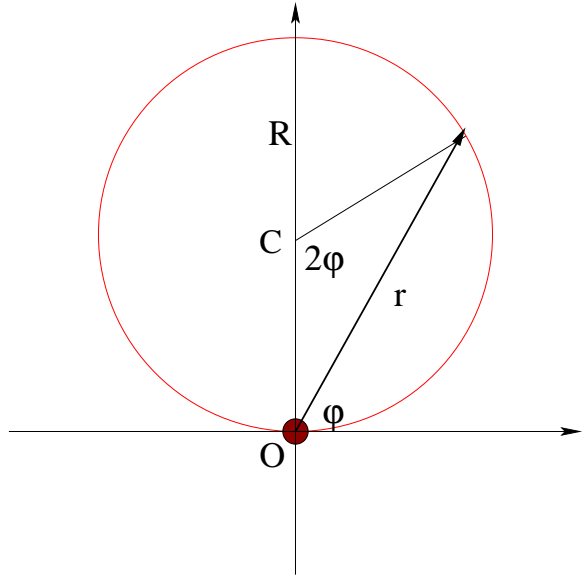}} \caption{\sl
Trajectory of the probe with zero angular momenta, passing through
co-moving observer located at O. The probe traces the Larmor
orbit, whose radius is $R$ and gyration speed is $\sqrt{E^2 -
m^2}$.}\label{trajectory}
\end{figure}
With the initial condition $r_1(0) = r_2(0) = 0$ that the probe
starts from co-moving observer's location $O$, the solution is
\be
 \phi_1(\tau) &=& \mu (E + q) \tau + \phi_{1}^{(0)} ~, \nonumber \\
 \phi_2(\tau) &=& \mu (E + q) \tau + \phi_2^{(0)} ~, \nonumber \\
 r_1(\tau) &=& \frac{\sqrt{E^2 - m^2}}{ |\mu(E + q)|}
 \sin \left(|\mu (E + q)| \tau \right) \cos \theta ~, \nonumber \\
 r_2(\tau) &=&  \frac{\sqrt{E^2 - m^2}}{ |\mu(E + q)|}
 \sin \left(|\mu (E + q)| \tau \right) \sin \theta ~, \nonumber \\
 t(\tau) &=& \left( E - \frac12 \left( \frac{E^2-m^2}{E + q}
 \right)\right) \tau +
 {1 \over 4 \mu} \frac{E^2-m^2}{(E + q)^2 }
 \sin  \left( 2 |\mu (E + q)| \tau \right) +t^{(0)} . \label{sol}
\ee
Here, $\theta$ ($0\le \theta \le \pi/2$) parametrizes projection
of the motion to each ${\bf R}^2$-plane, $\phi_1^{(0)},
\phi_2^{(0)}, t^{(0)}$ are parameters fixed by initial condition,
and the conserved energy $E$, Eq.(\ref{E}), ranges over $E \ge m$
or $E \le -m$ (to yield physically meaningful solution satisfying
Eq.(\ref{1stintegral})). In our convention, $E$, $m$ and $q$
carries mass dimension 1, and BPS conditions set inequalities $|E|
\geq m \geq |q|$. We thus see that probe's trajectory is
uni-directional: angular velocities $\dot\phi_1(\tau),
\dot\phi_2(\tau)$ revolve the same direction as the background
G\"odel universe Eq.(\ref{metric}) for sgn$(\mu(E+q))>0$ and
opposite direction for sgn$(\mu(E+q))<0$, respectively
\footnote{This feature is also reflected in the quantum dynamics
of the probes. See section 3.}. Under PT-conjugation of
Eq.(\ref{cpt}), the anti-probe follows the same trajectory but in
opposite directions.

The motion is reminiscent of the Landau problem mentioned earlier.
As evident from Eq.(\ref{sol}), probe's trajectory traces a
circular orbit on each ${\bf R}^2$-plane, as depicted in
Fig.\ref{trajectory}. One may describe the orbit with respect to
the center of the orbit $C$ (instead of co-moving observer's
location $O$). The gyration diameter $D=2R$, as depicted in
Fig.\ref{trajectory}, is set by the conserved energy
\be D = 2R= R_{\rm CTC} {\sqrt{E^2 - m^2} \over | E + q|},
\label{diameter} \ee
but is constant otherwise. Using trigonometry, gyration velocity
${\bf v}$ around $C$ (as measured in affine time) is also
determined as
\be {\bf v} = R \cdot 2\dot{\phi}_1 = R \cdot 2\dot{\phi}_2 = {\rm
sgn}(\mu (E + q)) \sqrt{E^2 - m^2}. \label{angularvelocity} \ee

\subsubsection{M-Graviton}
For a neutral probe, $q = 0$, corresponding to Kaluza-Klein modes
of the M-theory graviton, Eq.(\ref{sol}) reduces to the geodesics
studied in \cite{chandrasekhar, Goedel}. \hfill\break
$\bullet$ For the massless mode, $m=0$, the geodesics start out
radially from the origin $O$, and sweep out a circular gyration
orbit, touching the CTC horizon $R_{\rm CTC}$ associated with the
co-moving observer at $O$. That is,
\be r_1(\tau), \,\, r_2(\tau) \quad \le \quad R_{\rm CTC},
\nonumber \ee
as depicted as the leftmost circles in Fig.2. To reach the CTC
horizon, $\phi_1, \phi_2$ should sweep out $\pi/2$, so it takes
affine time $\Delta \tau = (\pi/2|\mu E|)$. Consequently, a
complete revolution around the orbit takes $\Delta \tau =
(\pi/|\mu E|)$ in affine time, and hence from Eq.(\ref{sol}),
$\Delta t = (\pi/2 |\mu|)$ in co-moving observer's time. Notice
that the gyration orbit is independent of graviton's energy $E$.
One can see this already from Eq.(\ref{1stintegral}) --- both the
harmonic frequency in the left-hand side and the constant of
motion in the right-hand side are proportional to $E^2$, and can
be rescaled away. \hfill\break
$\bullet$ For the massive modes, the geodesics extend radially
only up to the distance
\be r_1, \,\, r_2 \quad \le \quad R_{\rm CTC} \sqrt{1 - {m^2 \over
E^2} } \quad < \quad R_{\rm CTC}. \nonumber \ee
viz. the projected orbit has a radius smaller than the CTC horizon
$R_{\rm CTC}$ (Recall from Eq.(\ref{ctcsurface}) that the causal
region as perceived by the co-moving observer is separated from
the CTC region by the CTC horizon $r_{1,2} = R_{\rm CTC}$). It
follows that geodesics of M-theory graviton in the G\"odel
universe lie entirely within the causal region
\cite{chandrasekhar, Goedel} \footnote{Taking up this observation,
in \cite{Goedel}, it was proposed to cut off the region outside
the CTC horizon, paste it with flat space-time, and place the
holographic screen inside the causal region, where the area of the
holographic screen as prescribed in \cite{bakrey, bousso} is
maximized.}. Otherwise, the geodesics behave essentially the same
as the massless ones. In particular, it takes exactly the same
amount of affine and co-moving time for the gyration to undergo a
complete revolution.

\begin{figure}
\vskip-0.5cm \centerline{\includegraphics{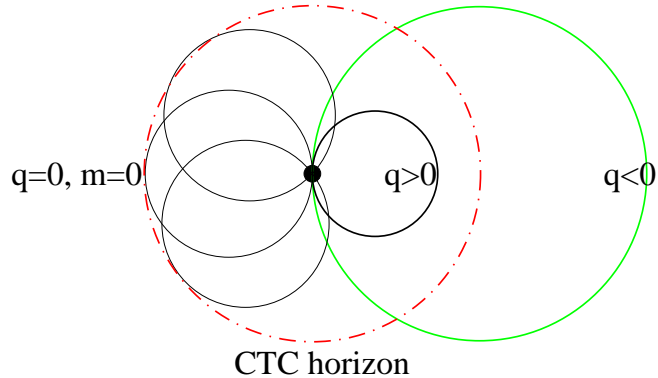}} \caption{\sl
Trajectories of various probes. The inner circle refers to the CTC
horizon as seen in the inertial frame of comoving observer at the
center, outside which closed time-like curves are present.
Geodesics of M-graviton, shown in the left for three different
angular initial conditions, are confined inside the CTC horizon.
Trajectory of M2-branes with $E>0$ are shown in the right for
$q>0$ and $q<0$, respectively. Notice that $q>0$ orbit is always
inside the CTC horizon, while $q<0$ orbit extends beyond the CTC
horizon.} \label{motions}
\end{figure}

\subsubsection{M2-branes}
For $q \ne 0$, corresponding to black M2-brane, the trajectory
Eq.(\ref{sol}) entails several new features. Because of the
background magnetic field, the M2-branes, being an electrically
charged probe, experiences the Lorentz force in addition to the
force exerted by the G\"odel universe rotation. Recalling that the
rotational force is uni-directional, depending on the sign of
M2-brane charge, the Lorentz force may act to add up to or to
cancel off the rotational force. It clearly points to the
possibility that the co-moving observer can arrange a probe brane
that travels beyond the CTC horizon and comes back. Let us analyze
how this may happen.

Eq.(\ref{sol}) indicates that the net effect depends crucially on
the relative sign between the energy $E$ and the charge $q$.
Consider an M2-brane with $E>m$, orbiting through co-moving
observer's location $O$. Diameter $D$ of the orbit is given by
Eq.(\ref{diameter}), so let us examine how $D$ depends on $E$ and
$q$. We are particularly interested in whether $D$ could be larger
than $R_{\rm CTC}$ for appropriate values of $E$ and $q$. We see
from Eq.(\ref{sol}) that, for BPS ${\rm M2}$-brane with positive
charge $q=m$,
\be D_{\rm sub} = R_{\rm CTC} \sqrt{(E- m)/(E+m)} < R_{\rm CTC}
\qquad {\rm for} \qquad {\rm sgn}(Eq) > 0 \, , \nonumber \ee
so the orbit gyrates entirely inside the CTC horizon of the
co-moving observer $O$. We will refer the trajectory as
"sub-gyration orbit" of diameter $D_{\rm sub}$. For BPS M2-brane
with negative charge $q=-m$,
\be D_{\rm super} = R_{\rm CTC}\sqrt{(E + m)/(E-m)} > R_{\rm CTC}
\qquad {\rm for} \qquad {\rm sgn}(Eq) < 0 \, , \nonumber \ee
so, as observed by the co-moving observer $O$, the orbit passes
through the CTC horizon and gyrate into the region where CTCs are
present! This is in stark contrast to the situation of M-graviton
geodesics studied in the previous section. We will refer the
trajectory as "super-gyration orbit" of diameter $D_{\rm super}$.
The two types of behavior are depicted as the rightmost two orbits
in Fig.\ref{motions}.

Information concerning M2-brane's dynamics for $E <-m$ are
obtainable by applying the PT conjugation Eq.(\ref{cpt}). One
finds that the two types of behavior are interchanged: with $E<
-m$, BPS M2-brane with positive charge $q = m$ traces
"super-gyration orbit" and extends beyond the CTC horizon, while
M2-brane with negative charge $q=-m$ traces "sub-gyration orbit"
and remains inside the CTC horizon.

Let us draw further comparison between "sub-gyration orbits" and
"super-gyration orbits". We consider the positive-energy branch
$E>m$ only, as the negative-energy branch $E<-m$ is obtainable by
PT conjugation of the positive-energy branch. \hfill\break
$\bullet$ Diameter of the sub-gyration orbit is a monotonically
{\sl increasing} function of $E$, ranging over $0 < D_{\rm sub} <
R_{\rm CTC}$. In contrast, diameter of the super-gyration orbit is
a monotonically {\sl decreasing} function of $E$, ranging over $
R_{\rm CTC} < D_{\rm super} < \infty$. Interestingly, in both
cases, orbits at extreme high-energy limit accumulate to the CTC
horizon, viz. the `velocity of light surface', and the gyration
velocity becomes infinite. In other words, in so far as dynamics
of M-theory probes is concerned, the CTC horizon defines a
universal infinite-momentum light-front frame, as gyration orbits
of both the M-graviton and M2-brane asymptote all to it as the
`velocity of light' surface. This suggests a viable {\sl
microscopic} holography of the G\"odel universe in M-theory
\cite{workinprogress}. \hfill\break
$\bullet$ From $t$-velocity Eq.(\ref{dott}), we see that, for
$L_{1,2}=0$ under consideration, the co-moving observer at $O$
would draw the standard interpretation of forward time-flow for $E
> m$ and backward time-flow for $E < -m$. Far away from $O$,
however, Eq.(\ref{dott}) indicates that the time-flow would be
seen reversed. Using the result Eq.(\ref{sol}), one finds that the
reversal takes place at a distance $r_{\rm r}$:
\be r_{\rm r} = R_{\rm CTC} \sqrt{E \over E + q}. \nonumber \ee
Comparing this with the diameter $D$ of the gyration orbit
Eq.(\ref{diameter}), we find that M2-branes whose energy and
charge have same sign would never reverse the time-flow, but
M2-branes whose energy and charge have opposite sign would do so.
At the moment of time-flow reversal, angular position $\phi_{1,2}$
of the probe M2-brane is at $\sin \phi_{\rm r} =
\sqrt{E/(E+|q|)}$.

\subsection{Probes with Nonzero Orbital Angular Momenta}
So far, we assumed that the probe carries no angular momentum. In
case the angular momenta $L_1, L_2$ are nonzero, new intriguing
features arise. \hfill\break
$\bullet$ Probe's orbits all migrate off the location of the
co-moving observer, and, if the orbital angular momentum meets a
suitable condition, penetrate into the CTC region. This is readily
seen as follows. Substitute Eq.\eqref{dotphi1}, \eqref{dotphi2}
and \eqref{dott} into \eqref{affine} with $e = 1$. We then obtain
\be (\dot r_1)^2 + (\dot r_2)^2  = (E^2-m^2) - V_{\rm
eff}(r_1,r_2) \nonumber \ee
where
\be V_{\rm eff} (r_1,r_2)= \left( \frac{L_1}{r_1} + \mu r_1 (E+q)
\right)^2  + \left( \frac{L_2}{r_2} + \mu r_2 (E+q) \right)^2 ~.
\label{r_1} \ee
The gyration orbits now range over $V_{\rm eff} (r_1,r_2) \le
(E^2-m^2)$, so
\be \left( \frac{L_1}{r_1} + \mu r_1 (E+q) \right)^2
&\le& (E^2 - m^2) \cos^2 \theta, \nonumber \\
\left( \frac{L_2}{r_2} + \mu r_2 (E+q) \right)^2 &\le& (E^2 - m^2)
\sin^2 \theta ~. \nonumber
\end{eqnarray}
Consider the positive-energy branch, $E>m$ \footnote{Again,
negative-energy branch, $E < -m$, is deducible by applying
PT-conjugation, under which $L_{1,2} \rightarrow - L_{1,2}$ as
well, to the following consideration.}. We see from the above
inequalities that not all angular momenta are physically allowed.
Those with $L_1, L_2 < 0$ are always possible, but those with
$L_1, L_2 > 0$ cannot be arbitrarily large. Rather, angular
momenta are bounded by
\be - \infty < (L_1, L_2) < L_{\rm max} (\cos \theta, \, \sin
\theta) \qquad {\rm where} \qquad L_{\rm max} \equiv {1 \over 4}
R_{\rm CTC} (E - q). \label{maxL} \ee
In this case, the classical turning points are
\be r_1^{(\pm)} &=& {D \over 2}\left\vert \cos \theta \pm
\sqrt{\cos^2 \theta -{L_1 \over L_{\rm max} } } \, \right\vert \nonumber \\
 r_2^{(\pm)} &=& {D \over 2}
 \left\vert \sin \theta \pm \, \sqrt{\sin^2 \theta - {L_2 \over
 L_{\rm max} } } \, \right\vert. \label{results} \ee
$\bullet$ Intuitively, one expects that orbital angular momentum
affects the orbits similar to the Lorentz force. The actual effect
is rather interesting. For $L_1, L_2 >0$, the gyration center
remains unchanged from that for zero angular momentum, viz.
located when projected on each ${\bf R}^2$-plane at distance ${1
\over 2} D \cos \theta$ and ${1 \over 2} D \sin \theta$,
respectively. On the other hand, the diameter of each orbit is
affected. On each ${\bf R}^2$-plane, the diameter is $D
\sqrt{\cos^2\theta - (L_1 /L_{\rm max})}$ and $D \sqrt{\sin^2
\theta - (L_2/L_{\rm max})}$, respectively. So, the gyration
orbits shrink once the orbital angular momentum is applied to the
probe {\sl same} direction as the rotation of the G\"odel universe
background. At and above the critical angular momenta
Eq.(\ref{maxL}), the orbits cease to exist. For $L_1, L_2 <0$, the
effects are reversed: the orbit diameter remains unchanged from
that for zero orbital angular momenta, viz. projected on each
${\bf R}^2$-plane, it is $D \cos \theta$ and $D \sin \theta$,
respectively, but the gyration center is shifted further away from
the co-moving observer $O$. So, once the orbital angular momentum
is applied to the probe the {\sl opposite} direction as the
rotation of the G\"odel universe background, the orbit slides off
from the hand of the co-moving observer. In particular, as
perceived by the co-moving observer at ${\cal O}$, the orbit
passes through the CTC horizon and travels the CTC region.
\hfill\break
$\bullet$ The time-flow $\dot{t}$ is now modified further by the
`spin-orbit coupling', the last term in Eq.(\ref{dott}), and can
be reverted, similar to the situation with the Lorentz force. This
is evident for $L_1, L_2 < 0$: if magnitude of $L_1, L_2$ is large
enough, gyration center migrates outside the CTC horizon, causing
the second term in Eq.(\ref{dott}) to dominate over the other
terms. For $L_1, L_2>0$, however, the coupling cannot revert the
time-flow due in part to the limit Eq.(\ref{maxL}) and the fact
that the orbit stays inside the CTC horizon.

\subsection{IIB String Theory Setup}
Instead of uplifting to M-theory, equivalently, to Type IIA string
theory, one can uplift the five-dimensional G\"odel universe
${\cal G}_5$ to Type IIB string theory, as originally demonstrated
in \cite{herdeiro}. The Type IIB background is given by
\be
 \d s^2_{\rm IIB} &=& g_{mn} \d x^{m} \d x^{n} +
 \left(\d y + \frac{1}{2\sqrt{3}} A_{m} \d x^{m}\right)^2 + \d s^2({X_4}) ,
 \label{iibmetric} \\
 H_3 &=& \d C_2 = \frac{1}{2\sqrt{3}} \d A \wedge
  \left(\d y + \frac{1}{2\sqrt{3}} A \right)
   - \frac{1}{2\sqrt{3}} {}^* \d A. \label{iibflux}
\ee
Here, the ten-dimensional coordinates are represented as
$(x^{m},y,z^a)$ with $m=0,\cdots,4$ and $a=1,\cdots,4$. The metric
$g_{mn}$ and the gauge connection $A$ are those of maximally
super-symmetric, five-dimensional G\"odel universe. The Hodge dual
${}^*$ is defined with respect to the five-dimensional metric
$g_{mn}$.

For the M-theory uplift, the probe that can couple to the
background magnetic flux in the five-dimensional G\"odel universe
was the M2-brane wrapped on two-cycle $\Sigma_2$ in the
six-dimensional `internal' space $X_6$. For the Type IIB string
theory uplift, a viable probe that can couple to the background
3-form flux Eq.(\ref{iibflux}) in the five-dimensional G\"odel
universe is the Kaluza-Klein mode along the 6-th dimension (whose
coordinates are labelled as $y$ in Eq.(\ref{iibmetric})).

Consider null geodesics in 6-dimensional G\"odel universe
Eq.(\ref{iibmetric}). Repeating the same analysis as in section
2.1, one finds the conserved, first integrals of motion:
\be
  E &=& \dot t + \mu (r_1^2 \dot \phi_1 +r^2_2 \dot \phi_2)  \qquad
  P_y = -\dot y - \mu (r_1^2 \dot \phi_1 +r^2_2 \dot \phi_2 )  \nonumber \\
  L_1 &=& r^2_1 \dot \phi_1 - \mu r^2_1 (\dot t - \dot y)  \quad \qquad
  L_2 = r^2_2 \dot \phi_2 - \mu r^2_2 (\dot t - \dot y) .
  \nonumber
\ee
%
%Using the gauge fixing condition for massless particle $g_{\mu
%\nu} \dot X^{\mu} \dot X^{\nu} = 0$ and the equations of motion,
%we can find the explicit solutions for the case without angular
%momenta $P_{\phi_{1,2}}=0$ as
%\begin{equation}\begin{split}
% \phi_1(\tau) &= \mu (E - P_y) \tau + \phi_{1}^{(0)} \\
% \phi_2(\tau) &= \mu (E - P_y) \tau + \phi_2^{(0)} \\
% r_1(\tau) &= {1 \over \mu} \frac{\sqrt{E^2 - P_y^2}}{| E - P_y |}
%   \sin \left(\mu (E - P_y) \tau \right) \cos \theta  \\
% r_2(\tau) &= {1 \over \mu} \frac{\sqrt{E^2 - P_y^2}}{| E - P_y |}
%   \sin \left(\mu (E - P_y) \tau \right) \sin \theta \\
% t(\tau) &= \left( E - \frac12 \left( \frac{E^2-P_y^2}{E - P_y}
% \right)\right) \tau +
% {1 \over 4 \mu} \frac{E^2-P_y^2}{(E - P_y)^2 }
%  \sin  \left( 2 \mu (E - P_y) \tau \right) +t^{(0)} \\
% y(\tau) &= \left( P_y - \frac12 \left( \frac{E^2-P_y^2}{E - P_y}
% \right)\right) \tau +
% {1 \over 4 \mu} \frac{E^2-P_y^2}{(E - P_y)^2 }
%  \sin  \left( 2 \mu (E - P_y) \tau \right) +y^{(0)} .
% \end{split} \end{equation}
One then finds that solution of the null geodesics is precisely
the same as Eq.(\ref{sol}) {\sl provided} the conserved charge
$P_y$ is equated with the electric charge $q$ of the wrapped
M2-brane.

\section{Quantum Dynamics} In the
previous section, we have seen that classical motions of the
probes, for both M-graviton and M2-brane, all trace gyration
orbits, much similar to the Landau problem. It thus brings up an
issue whether, quantum mechanically, probe's excitation energy
spectrum would exhibit discrete spectrum analogous to
Eq.(\ref{larmorfreq}). In this section, we will find that the
expectation is met precisely. We show that the energy spectrum is
precisely the same as that of two copies of the two-dimensional
simple harmonic oscillators, each associated with the two rotating
${\bf R}^2$-planes. In doing so, we will pay particular attention
to the energy spectrum of the `super-gyration orbits', which in
classical analysis encompassed outside the CTC horizon. By
comparing the energy spectrum with those for geodesics or
`sub-gyration orbit', one would hope to learn, if any, pathologies
associated with the CTCs. Our result shows that both the energy
spectrum and wave function of all probes are self-similar (Some
earlier works on wave dynamics in G\"odel universe include
\cite{KGeqn}).

\subsection{Graviton}
We begin with the M-theory graviton. Taking, for simplicity, the
polarization entirely along $X_6$ direction, the field equation of
a 5-dimensional graviton $\Phi$ is given by the massless
Klein-Gordon equation in the background Eq.(\ref{metric})
\be
 \square_5 \Phi = {1 \over \sqrt{-g_5}} \partial_m \left( g^{mn}_5
 \sqrt{-g_5} \partial_n \right) \Phi = 0. \label{KG}
 \ee

 %(-1 + \beta^2 (r^2_1 + r^2_2)) \partial_t \partial_t \Phi
 % - 2 \beta \partial_t \partial_{\phi_1} \Phi
 % - 2 \beta \partial_t \partial_{\phi_2} \Phi \\
 % &+ \frac{1}{r^2_1} \partial_{\phi_1}\partial_{\phi_1} \Phi
 % + \frac{1}{r^2_2} \partial_{\phi_2}\partial_{\phi_2} \Phi
 % + \partial_{r_1} \partial_{r_1} \Phi
 % + \partial_{r_2} \partial_{r_2} \Phi
 % + \frac{1}{r_1} \partial_{r_1} \Phi
 % + \frac{1}{r_2} \partial_{r_2} \Phi = 0 ~.
 %\label{KG}
As the G\"odel metric depends only on $r_1$ and $r_2$, it is
possible to decompose the field $\Phi$ into basis $\chi$'s via
separation of variables as
\be
 \Phi = \left\{\chi, \chi^*\right\} \qquad {\rm where} \qquad
 \chi = {\cal N} u_1 (r_1)u_2 (r_2)
 \exp ( i L_1 \phi_1 + i L_2 \phi_2 - i E t) ~.
 \label{separation}
\ee
where ${\cal N}$ is a normalization factor. The polar coordinates
$\phi_{1,2}$ range over $[0, 2 \pi]$, so the angular momenta $L_1,
L_2$ are quantized to integer units. The Klein-Gordon equation
Eq.(\ref{KG}) is then reduced to two coupled equations:
%
%\begin{equation}
%\begin{split}
% \Bigl[ (1 - \beta^2 (r^2_1 + r^2_2))E^2
%  - 2 \beta EP_1 &- 2 \beta EP_2
%  - \frac{P^2_1}{r^2_1}  - \frac{P^2_2}{r^2_2} \\ &
%  + \partial_{r_1} \partial_{r_1} + \partial_{r_2} \partial_{r_2}
%  + \frac{1}{r_1}\partial_{r_1}  + \frac{1}{r_2} \partial_{r_2}
%  \Bigr] \chi = 0 ~,
%\label{KG2}
%\end{split}
%\end{equation}
\be \left[ -\partial_{r_1}^2  - \frac{1 }{r_1}\partial_{r_1}
  + \left( \mu E r_1
  + \frac{L_1}{r_1} \right)^2 - E^2 \cos^2 \theta \right] u_1 (r_1) &=& 0 ~, \nonumber \\
 \left[ - \partial_{r_2}^2 - \frac{1 }{r_2}  \partial_{r_2}
  + \left( \mu E r_2 + \frac{L_2}{r_2} \right)^2 - E^2 \sin^2 \theta \right]
   u_2 (r_2) &=& 0  ~. \label{u1u2}
\ee
Here, $0 \le \theta \le \pi/2$ is a parameter introduced for
separation of the variables $r_1, r_2$. Eq.(\ref{u1u2}) takes the
form of Schr\"odinger equation of two-dimensional simple harmonic
oscillators whose natural frequency is given by $|\mu E|$ and
energy by $\{E^2 \cos^2 \theta - 2 \mu E L_1 \}$ or $\{E^2 \sin^2
\theta - 2 \mu E L_2 \}$, respectively. Normalizable solutions are
given in terms of the associated Laguerre polynomial ${\cal
L}_{\bf n}^\alpha$:
\be
 u_1 (r_1) &= r_1^{|L_1|} \exp \left( -
 \frac{1}{2} \vert \mu E \vert
 r^2_1\right)
{\cal L}_{\bf n_1}^{(|L_1|)} ( \vert\mu E \vert r^2_1)
%F \left( -n_1 , 1 + |P_{\phi_1}| ;  \beta E r^2_1 \right)
~, \nonumber \\
 u_2 (r_2) &= r_2^{|L_2|} \exp \left( -
 \frac{1}{2}\vert\mu E\vert
 r^2_2\right) {\cal L}_{\bf n_2}^{(|L_2|)} ( \vert\mu E\vert r^2_2)
%F \left( -n_2 , 1 + |P_{\phi_2}| ;  \beta E r^2_2 \right)
~, \label{neutralsol} \ee
where $\bf n_1, n_2$ are non-negative integers related to other
quantum numbers as
\be
 {\bf n_1} &=& - {1 \over 2} (1 + |L_1|) + \frac{1}{4 |\mu E|} \left[
 E^2 \cos^2 \theta - 2  \mu E L_1 \right] ~,\nonumber \\
 \nonumber \\
 {\bf n_2} &=& - {1 \over 2} (1 +|L_2|) + \frac{1}{4 |\mu E|} \left[
 E^2 \sin^2 \theta - 2 \mu E L_2 \right] ~.
 \label{gravitonns}
\ee

\begin{figure}
\vskip-1cm \centerline{\includegraphics{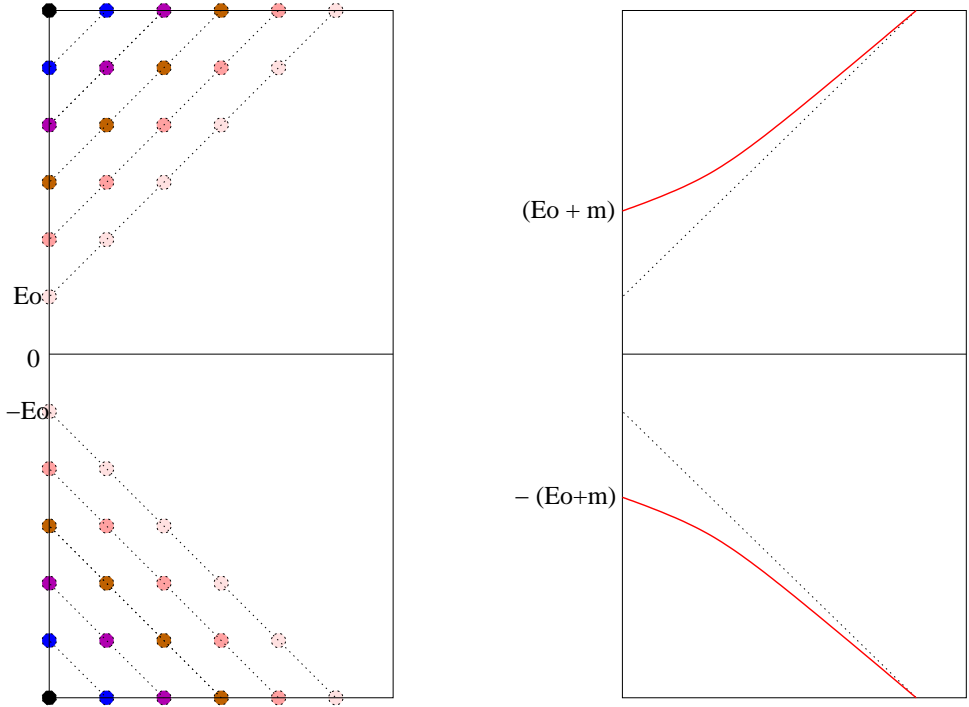}} \caption{\sl
Energy spectrum of massless (left) and massive (right) M-theory
gravitons. Horizontal axis refer to different multiplicities
associated with ${\bf n_1, n_2, m_1, m_2}$. The energy gap is
denoted as $E_0 = 4 \vert \mu \vert$.  } \label{spect}
\end{figure}

Adding the two radial quantum number relations in
Eq.(\ref{gravitonns}), we get
\be E^2 = 4 \vert\mu E \vert \Big({\bf n_1} + {\bf n_2} + 1 \Big)
+ 2 \Big( \mu E L_1 + \vert \mu E L_1 \vert \Big) + 2 \Big( \mu E
L_2 + \vert \mu E L_2 \vert \Big). \nonumber \ee
From the relation, we find the quantum energy spectrum $E$ of the
M-theory graviton in a remarkably simple analytic form:
\be |E| = 4 |\mu| {\cal N}_0 \label{spectrum} \ee
where
\be
 {\cal N}_0 = \left( {\bf n_1} +  {\bf n_2} + 1
 + {1 \over 2} \left\{|L_1| + {\rm sgn}(E\mu) \, L_1 \right\}
 + {1 \over 2} \left\{ |L_2| + {\rm sgn}(E\mu) \, L_2 \right\} \right)
 ~.
 \nonumber
\ee
Thus, positive and negative branches of the energy spectrum are
labelled by four non-negative integers $\bf n_1, n_2, m_1, m_2$
\footnote{The spectrum Eq.(\ref{gravspec}) is in agreement with
the supergravity modes of the closed string spectrum in G\"odel
universe \cite{harmark}.}
\be E^{(+)} &=& + 4 \vert \mu \vert \Big( {\bf n_1} + {\bf n_2}
+ {\bf m_1}^{(+)} + {\bf m_2}^{(+)} + 1 \Big) ~, \nonumber \\
E^{(-)} &=& - 4 \vert \mu \vert \Big( {\bf n_1} + {\bf n_2} + {\bf
m_1}^{(-)} + {\bf m_2}^{(-)} + 1 \Big) \, , \label{gravspec}\ee
where ${\bf m_{1,2}}^{(+)} = (|L_{1,2}| + {\rm sgn}(\mu)
L_{1,2})/2$ and ${\bf m_{1,2}}^{(-)} = ( |L_{1,2}| - {\rm
sgn}(\mu) L_{1,2})/2$, respectively. Notice that the two branches
Eq.(\ref{gravspec}) of the graviton spectrum are related each
other by the PT conjugation in Eq.(\ref{cpt}). The energy spectrum
and degeneracy is depicted in Fig. \ref{spect}.

A few remarks are in order.\hfill\break
\hfill\break $\bullet$ For massive M-graviton with mass $m$, the
energy spectrum is obtainable analogously in terms of ${\cal N}_0$
in Eq.(\ref{spectrum}) as:
\be E^{(\pm)} = \pm 2 \vert \mu \vert \left[ {\cal N}_0 + \left(
{\cal N}^2_0 + {1 \over 4} {m^2 \over \mu^2} \right)^{1/2}
\right]. \nonumber \ee
In Fig.\ref{spect}, we depict modification of the M-graviton
spectrum due to non-zero mass. Notice that the energy spectrum is
governed by the same orbital angular momentum quantum numbers
${\bf m_1}^{(\pm)}, {\bf m_2}^{(\pm)}$ as the massless one.
\hfill\break
$\bullet$ Recall that geodesics of M-graviton are closed orbits
around the CTC horizon, and never travel through the CTCs. The
spectrum Eq.(\ref{spectrum}) originates from Bohr-Sommerfeld
quantization of the M-graviton wave function around the classical
gyration orbit studied in section 2, so it is the counterpart of
the Landau level spectrum Eq.(\ref{larmorfreq}). Orbital angular
momentum of the M-graviton does not alter the interpretation, as
$L_1, L_2$ are conserved quantum numbers and renders the
M-graviton wave equation completely separable. Notice that
M-graviton's radial wave function, which determines the spatial
spread-out of the M-graviton, is symmetric under reversal of the
orbital angular momenta, $L_{1,2} \rightarrow - L_{1,2}$, while
the energy spectrum Eq.(\ref{gravspec}) is not. \hfill\break
$\bullet$ The ${\bf n_1}, {\bf n_2}$ quantum numbers are
interpretable as labelling `excitation' above the BPS ground state
of orbiting M-graviton. The zero-point energy proportional to $4
|\mu|$ in Eq.(\ref{gravspec}) may be attributed to the fact that
the null orbit is {\sl closed} on each ${\bf R}^2$-plane. The
frequency $|\mu|$ in Eq.(\ref{gravspec}) is the parameter that
sets the rotation and, in turn, the CTC horizon in the G\"odel
universe, so it is the counterpart of the Larmor frequency
$\Omega_{\rm L}$ in Eq.(\ref{larmorfreq}). Apparently,
quantization of the energy spectrum has nothing to do with the
presence of CNCs and CTCs in G\"odel universe. \hfill\break
$\bullet$ The angular momentum quantum numbers $L_1, L_2$ run over
all integer values, both positive and negative. Yet, the spectrum
Eq.(\ref{gravspec}) is affected only when the orbital angular
momentum quantum numbers take positive (negative) values,
respectively, for positive (negative) energy. It implies that, for
positive (negative) energy and negative (positive) orbital angular
momenta, the spectrum Eq.(\ref{gravspec}) is infinitely
degenerate. This is a direct reflection of orbit's classical
behavior with nonzero orbital angular momenta. Consider a
classical orbit with positive (negative) energy. As explained in
section 2.3, if positive (negative) orbital angular momenta are
applied to the M-graviton, orbit's diameter shrinks accordingly.
It implies that, by Bohr-Sommerfeld quantization rule,
quantum-mechanical energy ought to depend on the orbital angular
momentum quantum numbers. For a fixed `excitation' energy $E$ of
the M-graviton, the orbital angular momentum quantum numbers
cannot exceed the maximum
\be {\rm max} ({\bf m_1}, {\bf m_2}) = {E \over 4 |\mu|},
\nonumber \ee
and this is precisely the same upper bound $L_{\rm max}$ as the
classical counterpart Eq.(\ref{maxL}) (for $q=0$). Instead, if
negative (positive) orbital angular momenta are applied to the
M-graviton, orbit's diameter remains the same. By Bohr-Sommerfeld
quantization rule, it implies that quantum-mechanical energy ought
not to depend on the orbital angular momentum quantum numbers.
Rather, it renders the energy level degeneracy infinite,
reflecting the fact that orbit's gyration center can migrate off
all over each ${\bf R}^2$-plane once the orbit angular momenta are
cranked up arbitrarily high.

\subsection{M2-Brane}
Let us turn to the dimensionally reduced M2-brane. Interaction of
the M2-brane with the background gauge field $G_2 = \d C_1$ is
facilitated by the minimal coupling prescription $\partial_m
\rightarrow \partial_{m} + i (q/2 \sqrt{3}) C_{m}$ in
Eq.(\ref{KG}). In the Landau gauge Eq.(\ref{landaugauge}),
\be
 C_{\phi_1}(r_1) = 2 \sqrt{3} \mu r^2_1 \qquad {\rm and} \qquad
 C_{\phi_2}(r_2) = 2 \sqrt{3} \mu r^2_2 ~, \nonumber
\ee
the charged and massive counterpart of Eqs.\eqref{u1u2} becomes
\be
 \left[  -\partial_{r_1}^2  - \frac{1}{r_1} \partial_{r_1}
   + \left( (E + q)  \mu r_1 + \frac{L_1}{r_1} \right)^2
   - (E^2 - m^2)\cos^2 \theta
  \right] u_1 (r_1) &=& 0 \nonumber \\
 \left[- \partial_{r_2}^2 - \frac{1}{r_2} \partial_{r_2}
  + \left( (E + q) \mu r_2 + \frac{L_2}{r_2} \right)^2
  - (E^2 - m^2)\sin^2 \theta
   \right] u_2 (r_2) &=& 0  ~, \nonumber
 \ee
The solution is again obtainable in terms of the associated
Laguerre polynomials as
\be
 u_1 (r_1) &=&  r_1^{|L_1|}
 \exp \left( - {1 \over 2}|\mu (E + q)| r^2_1\right)
 {\cal L}_{\bf n_1}^{(|L_1|)} ( | \mu (E + q)| r^2_1)
% F \left( -n_1 , 1 + |P_{\phi_1}| ; \beta |E - 2 \sqrt3 e| r^2_1 \right)
~, \nonumber \\
 u_2 (r_2) &=& r_2^{|L_2|}
 \exp \left( - {1 \over 2} |\mu (E + q)| r^2_2\right)
{\cal L}_{\bf n_2}^{(|L_2|)} ( |\mu (E + q)| r^2_2)
% F \left( -n_2 , 1 + |P_{\phi_2}| ; \beta | E - 2 \sqrt3 e |  r^2_2 \right)
~, \label{chargedsol} \ee
with non-negative integer quantum numbers
\be
 {\bf n_1} &=& - \frac12 (1 + |L_1|)
 + \frac{1}{4 |\mu (E + q)|} \left( (E^2 - m^2) \cos^2 \theta  - 2
 \mu L_1 (E + q) \right)  ~, \nonumber\\
 {\bf n_2} &=& - \frac12 (1 + |L_2|) + \frac{1}{4 |\mu (E + q)|}
 \left( (E^2 - m^2) \sin^2 \theta - 2 \mu L_2 (E + q) \right) ~.
\nonumber \ee
It then follows that the dispersion relation is given by
\be E^2 - m^2 =  4 |\mu ( E + q )| {\cal N}_2\, ,  \ee
\label{energy}
where
\be {\cal N}_2 \equiv
 \left({\bf n_1} +  {\bf n_2} + 1 +
 \frac12 \left\{ |L_1| +  {\rm sgn} (\mu (E+q)) L_1 \right\}
 +\frac12 \left\{ |L_2| + {\rm sgn} (\mu (E+q)) L_2 \right\}
 \right)~. \label{n2} \ee
The right-hand-side of Eq.(\ref{energy}) is positive-definite, and
yields the two branches of the energy spectrum
\be E^{(+)}: \quad E > m \qquad {\rm and} \qquad  E^{(-)}: \quad E
< - m. \label{m2branch} \ee
For BPS M2-brane with $q = \pm m$, Eq.(\ref{energy}) is reduced to
$E = q \pm 4 |\mu| {\cal N}$. Keeping again track of sign of the
orbital angular momentum quantum numbers, we find that the energy
spectrum is given by
\be  E^{(+)} &=& q + 4 |\mu| \Big[{\bf n_1} + {\bf n_2} + {\bf
m_1}^{(+)} + {\bf m_2}^{(+)} + 1 \Big] \, , \nonumber \\
E^{(-)} &=& q - 4 |\mu| \Big[{\bf n_1} + {\bf n_2} + {\bf
m_1}^{(-)} + {\bf m_2}^{(-)} + 1 \Big] \, , \label{m2spec}\ee
subject to the condition Eq.(\ref{m2branch}).  Here, ${\bf
m}^{(\pm)}_{1,2} \equiv (\vert L_{1,2} \vert \pm {\rm sgn}(\mu)
L_{1,2})/2 = 0, 1, 2, \cdots$. Remarkably, apart from the rest
mass of the M2-brane, the resulting energy spectrum is identical
to that of the massless M-graviton studied in the previous
subsection. Notice that the energy spectra, $E^{(+)}$ and
$E^{(-)}$, are mapped to each other under the PT conjugation
Eq.(\ref{cpt}).

Several remarks are in order. \hfill\break
$\bullet$ As for the M-graviton, the energy spectrum
Eq.(\ref{m2spec}) depends sharply on the relative sign between the
energy and the orbital angular momenta. With positive (negative)
orbital angular momenta of M2-brane, orbit's diameter shrinks, so
the energy spectrum depends on the quantum numbers, ${\bf m_1},
{\bf m_2}$. For a fixed `excitation' energy $E$ of the M2-brane,
these quantum numbers cannot exceed the maximum
\be {\rm max} ({\bf m_1}, {\bf m_2}) = {(E-q) \over 4 |\mu|},
\nonumber \ee
and this is precisely the upper bound $L_{\rm max}$ of the
classical counterpart Eq.(\ref{maxL}) (for $q \ne 0$). With
negative (positive) orbital angular momenta of the M2-brane,
classical orbit's diameter remained unaffected, and the energy
spectrum is independent of ${\bf m_1}, {\bf m_2}$ quantum numbers.
Instead, the energy levels become infinitely degenerate,
reflecting classical result that orbit's gyration center can
migrate off all over each ${\bf R}^2$-plane once M2-brane's orbit
angular momenta are cranked up arbitrarily high. \hfill\break
$\bullet$ Classically, we learned in section 2 that, unlike
M-graviton, M2-brane probe can pass through the CTC horizon and
gyrate around in the region where the co-moving observer at $O$
perceives CTCs, even for zero orbital angular momenta. Quantum
mechanically, wave function and hence probability density of the
two probes Eqs.(\ref{neutralsol}) and Eqs.(\ref{chargedsol}),
respectively, are self-similar. The two are merely related each
other by (energy-dependent) rescaling radial coordinates by an
energy-dependent factor $\sqrt{1 + (q/E)}$. Moreover, for both,
the energy scale of the excitation spectrum is set by one and the
same Larmor frequency, $4 |\mu|$. It may be that, in order for the
quantum mechanics to reveal full-fledged pathologies of the CTCs
present, more refined dynamical processes, e.g. interference of
multi-body wave functions or nonlinear mode-mode interactions,
should be considered. \hfill\break
$\bullet$ Recall that, in the foregoing analysis, we have tacitly
assumed that the orientation of the two-cycles M2-brane wraps on
is such that the M2- and $\overline{\rm M2}$-branes carry electric
charge $q$ positive and negative, respectively. Upon reversing
orientation of the two-cycles \footnote{ Orientation of the normal
directions needs to be reversed concurrently so that the net
orientation of $X_6$ is unaffected.}, the sign of the electric
charge $q$ would flip. All the foregoing analysis would be the
same except that $q$ should be taken the opposite. Still, as $E
> m$ and $E < - m$, the spectrum is isomorphic to the above.

\subsection{IIB String Theory Setup}
By taking similar steps, one can find excitation energy spectrum
of IIB-graviton probe by solving the six-dimensional wave
equation, $\square_6 \Phi = 0$, in the background
Eq.(\ref{iibmetric}). Making use of the conserved quantum numbers
identified in section 2.4, we again take separation of variables
as
\be  \Phi = (\chi, \chi^*) \qquad \chi = {\cal N} u_1 (r_1)u_2
(r_2)
 \exp ( i L_1 \phi_1 + i L_2 \phi_2 + i P_y y - i E t) . \nonumber
\ee
Radial parts of the wave equation are then given by
\be
 \left[  -\partial_{r_1}^2  - \frac{1}{r_1} \partial_{r_1}
   + \left( (E - P_y)  \mu r_1 + \frac{L_1}{r_1} \right)^2
  - (E^2 - P_y^2)\cos^2 \theta
  \right] u_1 (r_1) &=& 0 \\
 \left[-\partial_{r_2}^2 - \frac{1}{r_2} \partial_{r_2}
  -\left( (E - P_y) \mu r_2 + \frac{L_2}{r_2} \right)^2
  - (E^2 - P_y^2)\sin^2 \theta \right] u_2 (r_2) &=& 0 \, . \nonumber
\ee
One finds that they are the same as the ones for the M-graviton
and the M2-brane for $P_y = 0$ and $P_y = -q$, respectively, so
the wave function and energy spectrum are exactly the same as
those.

\section{Discussion}
In this paper, we have studied classical and quantum dynamics of
M-graviton and M2-brane probes in G\"odel universe, with
particular attention to possible effects of the CTCs present
beyond the CTC horizon of a co-moving observer. Our results are
summarized as follows. \hfill\break
$\bullet$ On each rotating ${\bf R}^2$-plane, all probes trace
gyration orbits, much similar to the Landau problem,
Eq.(\ref{larmorradius}). For each orbit, gyration center and
diameter depends on probe's conserved quantum numbers: energy $E$,
angular momenta $L$, and 3-form potential charge $q$. \hfill\break
$\bullet$ For zero orbital angular momenta, orbit of M-graviton
(geodesics) traces radial distance between a co-moving observer
located at origin $O$ and the CTC horizon $R_{\rm CTC}$ perceived
by the co-moving observer. Orbit of M2-brane traces between a
co-moving observer at $O$ and the maximal distance $D$ (diameter
of the orbit). For sgn$(Eq) >0$, $D$ is less than $R_{\rm CTC}$,
so as perceived by the co-moving observer, the orbit does not
travel through the region where CTCs are present. For sgn$(Eq)<0$,
$D$ is larger than $R_{\rm CTC}$, and the orbit travels through
the region where CTCs are present. \hfill\break
$\bullet$ Nonzero orbital angular momenta $L$ modify probe's orbit
as well. For sgn$\mu L >0$, the gyration takes place at the same
center as $L=0$ but the gyration diameter is shrunken. For sgn$\mu
L <0$, the gyration diameter is the same as $L=0$ but the center
migrates away from the co-moving observer. \hfill\break
$\bullet$ Quantum mechanically, probe's wave functions and
excitation energy spectrum are all self-similar, independent of
whether probe's classical orbit is larger or smaller than the CTC
horizon. Interestingly, under the reversal of the orbital angular
momentum, the wave functions are invariant.

Our results indicate that further investigation is imperative for
exploring full-fledged pathology of the G\"odel universe in
M-theory. It is evident that naive application of quantum field
theory approximation to M-theory on G\"odel universe is
ill-defined. For example, because G\"odel universe is not globally
hyperbolic, inner products is ill-defined. As such, expectation
value of the energy-momentum tensor $\langle T_{mn} \rangle$ and
back-reaction thereof are not computable.

An attitude one may take for the G\"odel universe is that it is
similar to negative-mass Schwarzschild or extremal
Reissner-Nordstrom black hole. Formally, the latter is a solution
with a naked time-like singularity, repelling all time-like
geodesics. For both negative-mass black holes and G\"odel
universe, they are not likely to be formed out of physical
processes in initial space-time without pathologies such as CTCs
or singularities. Supersymmetries preserved by the G\"odel
universe would not help much as the negative-mass extremal
Reissner-Nordstrom black hole as well is embeddable as a
supersymmetry preserving configuration. Our tentative attitude is
though sympathetic to \cite{Goedel}: until proven inconsistent,
the G\"odel universe in M-theory deserves further inspection,
including possible observational constraints \cite{obs}.

\subsection*{Acknowledgement}
We acknowledge Eric Gimon, Rajesh Gopakumar, Carlos Herdeiro, Gary
Horowitz, Oleg Lunin, Juan Maldacena, Fumihiko Sugino, Hiromitsu
Takayanagi and Tadashi Takayanagi for enlightening discussions.
This work was done while SJR was a Member at the Institute for
Advanced Study. He thanks the School of Natural Sciences for
hospitality and for the grant in aid from the Fund for Natural
Sciences.

\vskip1cm

\section*{Appendix}
\subsection*{Five-Dimensional G\"odel metric}
In the bipolar coordinates, nonzero components of the inverse
metric are
\be
 && g^{tt} = -1 + \mu^2 (r^2_1 + r^2_2) \qquad \qquad g^{r_1 r_1} = g^{r_2 r_2} =1 , \nonumber \\
 && g^{\phi_1 \phi_1} = \frac{1}{r^2_1} \quad\quad
 g^{\phi_2 \phi_2} = \frac{1}{r^2_2} \quad\quad  g^{\phi_1 t} = g^{\phi_2 t}
 = - \mu~. \nonumber
\ee
Nonzero components of the Christoffel connection are:
\be
\begin{array}{ccc}
\Gamma^t_{r_1 t} = \mu^2 r_1 \quad\,\, & \Gamma^{r_1}_{\phi_1 t} =
\mu r_1 \quad \quad \,\,\, & \Gamma^{\phi_1}_{r_1 t} = - \mu / r_1 \\
\Gamma^t_{\phi_1 r_1} = (\mu r_1)^3 & \quad \qquad
\Gamma^{r_1}_{\phi_1 \phi_1} = - r_1 ( 1 - 2 \mu^2 r_1^2) &
\quad\qquad \Gamma^{\phi_1}_{\phi_1 r_1} =  - {\mu^2 r_1} + 1/ r_1\\
\Gamma^t_{\phi_2 r_1} = \mu^3 r_1 r_2^2 & \Gamma^{r_1}_{\phi_1
\phi_2} = \mu^2 r_1 r_2^2 \,\,\,\, & \qquad
\Gamma^{\phi_1}_{\phi_2 r_1} = - \mu^2 r^2_2 / r_1\\
\Gamma^t_{r_2 t} = \mu^2 r_2 \quad\, & \Gamma^{r_2}_{\phi_2 t} =
\mu r_2 \quad \quad \,\,\, & \Gamma^{\phi_2}_{r_2 t} = - \mu / r_2\\
\Gamma^t_{\phi_2 r_2} = (\mu r_2)^3 & \qquad\quad
\Gamma^{r_2}_{\phi_2 \phi_2} = -r_2 ( 1 - 2 \mu^2 r_2^2) &
\qquad \quad\,\,\, \Gamma^{\phi_2}_{\phi_2 r_2} = - \mu^2 r_2 + 1/ r_2\\
\Gamma^t_{\phi_1 r_2} = \mu^3 r_1^2 r_2 & \Gamma^{r_2}_{\phi_1
\phi_2} = \mu^2 r_1^2 r_2 \,\,\,\, & \qquad
\Gamma^{\phi_2}_{\phi_1 r_2} = - \mu^2 r_1^2 / r_2
\end{array} \nonumber
\ee

\subsection*{Six-Dimensional G\"odel Universe}
For the six-dimensional G\"odel universe ${\cal G}_6$ in Type IIB
string theory, non-zero components of the inverse metric are
\be && g^{tt} = -1 + \mu^2 (r^2_1 + r^2_2) \quad g^{\phi_1 t} =
g^{\phi_2 t} =  g^{y\phi_1}=g^{y\phi_2} = - \mu \quad
    g^{yt}=\mu^2(r^2_1 + r^2_2) \, , \nonumber \\
   &&  g^{r_1 r_1} = g^{r_2 r_2} =1 \qquad
 g^{\phi_1 \phi_1} = \frac{1}{r^2_1} \qquad
 g^{\phi_2 \phi_2} = \frac{1}{r^2_2} \qquad
 g^{yy} = 1+\mu^2 (r^2_1 +r^2_2) \, , \nonumber
\ee
and non-zero components of the Christoffel connection are
\be && \Gamma^t_{r_1 t} = \Gamma^y_{r_1 t} = - \Gamma^y_{r_1 y} =
- \Gamma^t_{r_1 y} = \mu^2 r_1 \qquad
\Gamma^{r_1}_{\phi_1 t} = - \Gamma^{r_1}_{\phi_1 y} = \mu r_1 \nonumber \\
&& \Gamma^{r_1}_{\phi_1 \phi_1} = -r_1 \qquad
 \Gamma^{\phi_1}_{\phi_1 r_1} = \frac{1}{r_1} \qquad
 \Gamma^{\phi_1}_{r_1 t} = - \Gamma^{\phi_1}_{r_1 y} = - \frac{\mu}{r_1}
\nonumber \ee
 along with those replaced with $(1 \leftrightarrow 2)$.


\begin{thebibliography}{999}

\bibitem{godel}
K. G\"odel, {\em An example of a new type of cosmological
solutions of Einstein's field equations of gravitation}, Rev. Mod.
Phys. {\bf 21}, 447 (1949).

\bibitem{hawking} S.W. Hawking, {\em The chronology protection
conjecture}, Phys. Rev. D {\bf 46}, 603 (1992).

\bibitem{solutions} J.P. Gauntlett {\sl et.al.}, {\em All
supersymmetric solutions of minimal supergravity in
five-dimensions}, [arXiv:hep-th/0209114].

\bibitem{harmark} T.~Harmark and T.~Takayanagi, {\em Supersymmetric Goedel
universes in string theory}, [arXiv:hep-th/0301206].
%%CITATION = HEP-TH 0301206;%%

\bibitem{gibbons}
G.W. Gibbons and C. Herdeiro, {\em Supersymmetric rotating black
holes and causality violations}, Class. Quant. Grav. {\bf 16},
3619 (1999) [arXiv:hep-th/9906098].

\bibitem{herdeiro}
C.A. Herdeiro, {\em Special properties of five-dimensional BPS
rotating black holes}, Nucl. Phys. B{\bf 582}, 363 (2000)
[arXiv:hep-th/0003063]; ibid. {\em Spinning deformations of the
D1-D5 system and a geometric resolution of closed timelike curves}
[arXiv:hep-th/0212002].

\bibitem{dyson}
L. Dyson, {\em Chronology Protection in string theory}
[arXiv:hep-th/0302052].

\bibitem{leigh}
R. Biswas, E. Keski-Vakkuri, R.G. Leigh, S. Nowling and E. Sharpe,
{\em The taming of closed timelike curves} [arXiv:hep-th/0304241].

\bibitem{Goedel}
E.~K.~Boyda, S.~Ganguli, P.~Horava and U.~Varadarajan, {\em
Holographic protection of chronology in universes of the Goedel
type}, [arXiv:hep-th/0212087].
%%CITATION = HEP-TH 0212087;%%

\bibitem{bakrey} D. Bak and S.-J. Rey, {\em Cosmic Holography},
Class. Quant. Grav. {\bf 17}, L83 (2000) [hep-th/9902173].
%%CITATION = HEP-TH 9902173;%%

\bibitem{bousso} R. Bousso, {\em Holography in General
Space-Times}, JHEP {\bf 9906} (1999) 028 [arXiv:hep-th/9906022].

\bibitem{drukker} N. Drukker, B. Fiol and J. Sim\'on, {\em
G\"odel's Universe in a Supertube Shroud} [arXiv:hep-th/0306057].

\bibitem{chandrasekhar} S. Chandrasekhar and J.P. Wright, {\em
The geodesics in G\"odel universe}, Proc. Nat. Acad. Sci. {\bf
47}, 341 (1961).

%\bibitem{Charged}
%B.~D.~Figueiredo, {\em G\"odel-type spacetimes and the motions of
%charged particles}, Class.\ Quant.\ Grav.\  {\bf 15}, 3849 (1998).

\bibitem{workinprogress} Work in progress.

\bibitem{KGeqn}
B.~Mashhoon, {\em Influence of gravitation on the propagation of
electromagnetic radiation}, Phys.\ Rev.\ D {\bf 11}, 2679
(1975);\\
%%CITATION = PHRVA,D11,2679;%%
W.~A.~Hiscock, {\em Scalar perturbations in the Godel universe},
Phys.\ Rev.\ D {\bf 17}, 1497 (1978).
%%CITATION = PHRVA,D17,1497;%%

\bibitem{obs}
E.F. Bunn, P. Ferreira and J. Silk, {\em How anisotropic is our
universe?}, Phys. Rev. Lett. {\bf 77}, 2883 (1996)
[astro-ph/9605123].
%%CITATION = ASTRO-PH/9605123];%%


\end{thebibliography}
\end{document}